\documentclass[format=acmsmall, review=false, screen=true]{acmart}
\renewcommand\footnotetextcopyrightpermission[1]{}

\usepackage[utf8]{inputenc}
\usepackage{amsmath,amssymb,amsfonts}
\usepackage{listings}
\usepackage{seqsplit}
\usepackage{tabularx}
\usepackage{multirow}

\usepackage[ruled]{algorithm2e} 

\SetAlFnt{\small}
\SetAlCapFnt{\small}
\SetAlCapNameFnt{\small}
\SetAlCapHSkip{0pt}
\IncMargin{-\parindent}

\newcommand{\ie}{i.e.\xspace}

\lstdefinelanguage{JavaScript}{
  keywords={break, case, catch, continue, debugger, default, delete, do, else, false, finally, for, function, if, in, instanceof, new, null, return, switch, this, throw, true, try, typeof, var, void, while, with},
  morecomment=[l]{//},
  morecomment=[s]{/*}{*/},
  morestring=[b]',
  morestring=[b]",
  ndkeywords={class, export, boolean, throw, implements, import, this},
  keywordstyle=\color{blue}\bfseries,
  ndkeywordstyle=\color{darkgray}\bfseries,
  identifierstyle=\color{black},
  commentstyle=\color{purple}\ttfamily,
  stringstyle=\color{red}\ttfamily,
  sensitive=true
}

\acmJournal{TWEB}

\begin{document}

\title[Browser Fingerprinting: A survey]{Browser Fingerprinting: A survey}

\author{Pierre Laperdrix}
\affiliation{%
  \institution{CNRS, Univ Lille, Inria Lille}
  \city{Lille}
  \country{France}
}
\email{pierre.laperdrix@inria.fr}

\author{Nataliia Bielova}
\affiliation{%
  \institution{Inria Sophia Antipolis}
  \city{Sophia Antipolis}
  \country{France}
}
\email{nataliia.bielova@inria.fr}

\author{Benoit Baudry}
\affiliation{%
  \institution{KTH Royal Institute of Technology}
  \city{Stockholm}
  \country{Sweden}
}
\email{baudry@kth.se}

\author{Gildas Avoine}
\affiliation{%
  \institution{Univ Rennes, INSA Rennes, CNRS, IRISA}
  \city{Rennes}
  \country{France}
}
\email{gildas.avoine@irisa.fr}

\begin{abstract}
With this paper, we survey the research performed in the domain of browser fingerprinting, while providing an accessible entry point to newcomers in the field.
We explain how this technique works and where it stems from.
We analyze the related work in detail to understand the composition of modern fingerprints and see how this technique is currently used online.
We systematize existing defense solutions into different categories and detail the current challenges yet to overcome.
\end{abstract}


\maketitle

\makeatletter
\def\@journalNameShort{arXiv}
\makeatother

\section{Introduction}
\label{sec:introduction}

The web is a beautiful platform and browsers give us our entry point into it.
With the introduction of HTML5 and CSS3, the web has become richer and more dynamic than ever and it has now the foundations to support an incredible ecosystem of diverse devices from laptops to smartphones and tablets.
The diversity that is part of the modern web opened the door to device fingerprinting, a simple identification technique that can be used to collect a vast list of device characteristics on several layers of the system.
As its foundations are rooted into the origin of the web, browser fingerprinting cannot be fixed with a simple patch.
Clients and servers have been sharing device-specific information since the beginning to improve user experience.

The main concept behind browser fingerprinting is straight-forward: collecting device-specific information for purposes like identification or improved security.
However, when this concept is implemented, its exact contours are constantly changing as its mechanisms are entirely defined by current web browser technologies.

\subsection{Definition}
\label{subsec:definition}
A \textbf{browser fingerprint} is a set of information related to a user's device from the hardware to the operating system to the browser and its configuration.
\textbf{Browser fingerprinting} refers to the process of collecting information through a web browser to build a fingerprint of a device.
Via a simple script running inside a browser, a server can collect a wide variety of information from public interfaces called Application Programming Interface (API) and HTTP headers.
An API is an interface that provides an entry point to specific objects and functions.
While some APIs require a permission to be accessed like the microphone or the camera, most of them are freely accessible from any JavaScript script rendering the information collection trivial.
Contrarily to other identification techniques like cookies that rely on a unique identifier (ID) directly stored inside the browser, browser fingerprinting is qualified as completely \textit{stateless}.
It does not leave any trace as it does not require the storage of information inside the browser.

For the rest of this article, the terms ``browser fingerprint'' and ``device fingerprint'' will be used interchangeably.
We also consider cross-browser device fingerprinting as detailed in Section~\ref{subsubsec:cross}.
However, it should be noted that we focus only on information collected through a web browser.
We do not cover the identification of devices through smartphone applications like Kurtz et al.~\cite{kurtz2016fingerprinting} or Wu et al.~\cite{wu2016zero} as they have access to more information than with a simple browser and they require additional permissions to get installed.
We also do not focus on the analysis of the structure of network packets similar to the features offered by tools like nmap~\cite{nmap} as they fall out of context of what the browser has access to.
Finally, we do not study how the IP address or the geolocation of the user can contribute to the identification of a device.
While they can be used to complement a fingerprint, we focus here on what can be done entirely from the information given by a web browser.

\subsection{Contributions}
\label{subsec:contributions}
The goal of this work is twofold: first, to provide an accessible entry point for newcomers by systematizing existing work, and second, to form the foundations for future research in the domain by eliciting the current challenges yet to overcome.
We accomplish these goals with the following contributions:
\begin{itemize}
\item A thorough survey of the research conducted in the domain of browser fingerprinting with a summary 
of the framework used to evaluate 
the uniqueness of browser fingerprints and their adoption on the web.
\item An overview of how this technique is currently used in both research and industry.
\item A taxonomy that classifies existing defense mechanisms into different categories, providing a high-level view of the benefits and drawbacks of each of these techniques.
\item A discussion about the current state of browser fingerprinting and the challenges it is currently facing on the science, technological, business and legislative aspects.
\end{itemize}

\subsection{Organization}
\label{subsec:organization}
The remainder of this work is organized as follows.
Section~\ref{sec:history} reports on the evolution of web browsers over the years to gain an understanding of why browser fingerprinting became possible.
Section~\ref{sec:fingerprinting} provides a detailed survey on the work done in the domain.
Section~\ref{sec:defense} introduces the different approaches designed to protect users against it.
Section~\ref{sec:challenges} discusses the 
different usage of this technique and the current challenges in both research and industry.
Section~\ref{sec:conclusion} concludes this paper.
\section{A brief history of web browsers}
\label{sec:history}

In this section, we look at how web browsers turned from HTML renderers to full-fledged embedded operating systems.

\subsection{Indicating browser limitations with the user-agent header}
One of the key ideas of the early development of the web is that, for anyone to get access to this vast network of machines, it should be device agnostic, \ie run on any device with any type of architecture.
HTTP~\cite{rfcHTTP} and HTML~\cite{rfcHTML} were born from that need of having a universal way of communicating between machines and in the early 90s, web browsers started to appear from various teams around the world to support these soon-to-be standards.
However, as the foundations of the web started to evolve to keep pushing what is possible online, not every browser and not every platform supported the latest additions.
Some browsers conformed to only a subset of the specifications and developed their own features.
This started the now infamous era of the ``Best viewed with X'' banners.

To prevent incompatibility problems, the HTTP protocol includes the ``User-Agent request-header''~\cite{rfcHTTP}.
Browsers started to include their name, their version and even sometimes the platform on which they were running to avoid particular user agent limitations.
As reported by \cite{historyUA1} and \cite{historyUA2}, the story of the user-agent header is very rich and it keeps writing itself today as modern browsers still bear the legacy of the very first browsers.
The information contained in this header has become complex as browser vendors started copying the value of their competitors to declare their compatibility with a different rendering engine.
For example, the user-agent for version 68 of a Chrome browser running on Linux is the following:
\begin{lstlisting}[breaklines=true,breakatwhitespace=true]
Mozilla/5.0 (X11; Linux x86_64) AppleWebKit/537.36 (KHTML, like Gecko) Chrome/68.0.3440.75 Safari/537.36
\end{lstlisting}
The only relevant pieces of information here are ``(X11; Linux x86\_64)'' and ``Chrome/68.0.3440.75''.
Other strings like ``Gecko'', ``KHTML'' or ``Safari'' are present to declare their compatibility with other layout engines.
The string ``Mozilla/5.0'' even dates back from the time where the first ever version of Firefox was released to the public.
All modern web browsers now include it in the user-agent header for no particular reason.

In the end, the user-agent header can be considered as the very first piece of information that deliberately indicated differences between devices to help developers take into account browser limitations.

\subsection{Bridging the gap between web browsers and native software applications}
At the very beginning of the web, pages needed to be reloaded completely to allow live modifications.
In 1995, Brendan Eich added a scripting language called JavaScript inside the Netscape Navigator to make web pages more dynamic.
From then on, the language quickly gained a lot of traction and was implemented in most browsers in the months following its introduction.
The specification of the language became standardized in June 1997 under the name ``ECMAScript'', with JavaScript being the most well known of its implementations at the time.

As the language started growing and as browsers started to offer more and more features to their users, developers pushed to create a bridge between the browser and the platform it is running on.
The goal was to incorporate information from the user's environment inside the browser to feel more like a native software application.
The very first edition of the ECMAScript specification offers the first traces of such integration with details on the ``Date'' object~\cite{ecma1}.
To conform to the specification, browsers directly got the timezone of the device from the operating system.

\subsection{The development of modern APIs}
The modern browser has slowly shifted from being a tool that displays simple HTML pages to a very rich multimedia platform compatible with many formats and devices. 
Over the years, the W3C has developed many novel web standards to offer a rich browsing experience to users and to support the rising popularity of mobile browsing.

The Canvas API ``provides objects, methods, and properties to draw and manipulate graphics on a canvas drawing surface''~\cite{apiCanvas}.
Users can draw and animate any number of shapes and they can render textual content directly in the browser by using the graphical capabilities of the device. 
The Khronos Group~\cite{apiWebGL} introduced WebGL which is a graphics API that can render interactive 3D objects in the browser and manipulate them through JavaScript without the need for plugins. 
The Web Audio API provides an interface to create a pipeline to process audio~\cite{apiAudio}. 
By linking audio modules together, anyone can generate audio signals and apply very specific operations like compression or filtering to generate a very specific output.
Additional APIs were introduced to enrich the user experience like WebRTC for real-time communications~\cite{apiWebRTC}, Geolocation for real-time positioning~\cite{apiGeo} or WebAssembly to take browser performance to the next level~\cite{apiAssembly}.
Other APIs are still being designed and discussed like WebPayments~\cite{apiPayments} and WebXR~\cite{apiXR} to make the web an even richer platform.
Finally, modern browsers embraced user customization from the beginning by allowing anyone to develop their own browser extensions, small add-ons that can extend the functionalities of a browser.
The most popular ones today are ad blockers, password managers or video downloaders~\cite{firefoxExtension}.
\section{Browser fingerprinting}
\label{sec:fingerprinting}

In this section, we answer the fundamental questions surrounding the browser fingerprinting domain: Where does it stem from? How effective is it? How do we evaluate current techniques? How widespread is it on the web?
The aim of this section is to provide a thorough survey of the research conducted in the domain of browser fingerprinting with a summary of current techniques.

\subsection{Discovery of browser fingerprinting}
\label{subsec:pioneers}
In 2009, Mayer investigated if the differences that stem from the origins on the Internet could lead to the deanonymization of web clients~\cite{mayer2009}. 
Especially, he looked to see if differences in browsing environments could be exploited by a remote server to identify users.
He noticed that a browser could present ``quirkiness'' that came from the operating system, the hardware and the browser configuration.
He conducted an experiment where he collected the content of the \textit{navigator}, \textit{screen}, \textit{navigator.plugins}, and \textit{navigator.mimeTypes} objects of browsers who connected to the website of his experiment.
Out of 1328 clients, 1278 (96.23\%) could be uniquely identified. 
However, he added that the small scale of his study prevented him from drawing a more general conclusion.

A year later, Peter Eckersley from the Electronic Frontier Foundation (EFF) conducted the Panopticlick experiment~\cite{eckersley2010Pano}.
By communicating on social media and popular websites, he amassed 470,161 fingerprints in the span of two weeks.
Contrarily to Mayer, the amount of collected fingerprints gives a much more precise picture on the state of device diversity on the web.
With data from HTTP headers, JavaScript and plugins like Flash or Java, 83.6\% of fingerprints were unique.
If users had enabled Flash or Java, this number rose to 94.2\% as these plugins provided additional device information.
This study coined the term ``browser fingerprinting'' and was the first to prove that it was a reality on a very large scale.  
The privacy implications that emerged from it are really strong as a device with a not-so-common configuration can easily be identified on the Internet.

\subsection{Advancing fingerprinting}
\label{subsec:attributes}

\subsubsection{Full example}

Table~\ref{table:example} provides a full example of the main attributes collected in the browser fingerprinting literature along with their source.

\newcolumntype{M}[1]{>{\centering\arraybackslash}m{#1}}

\begin{table*}[h!]
\center
\caption{Example of a browser fingerprint.}
\label{table:example}

\scalebox{0.85}{
\begin{tabular}{ M{3cm} | M{1.6cm} | m{10cm} }
Attribute & Source & Example \\ \hline

	User agent & HTTP header & Mozilla/5.0 (X11; Linux x86\_64) AppleWebKit/537.36 (KHTML, like Gecko) Chrome/64.0.3282.119 Safari/537.36 \\ \hline
     Accept & HTTP header &
     \seqsplit{text/html,application/xhtml+xml,application/xml;q=0.9,image/webp,image/apng,*/*;q=0.8} \\ \hline
      Content encoding & HTTP header &
     gzip, deflate, br\\ \hline
      Content language & HTTP header &
     en-US,en;q=0.9\\ \hline
      List of plugins & JavaScript &
     Plugin 1: Chrome PDF Plugin. Plugin 2: Chrome PDF Viewer. Plugin 3: Native Client. Plugin 4: Shockwave Flash...\\  \hline
      Cookies enabled & JavaScript & yes \\ \hline
      Use of local/session storage & JavaScript & yes\\ \hline
      Timezone & JavaScript & -60 (UTC+1)\\ \hline
      Screen resolution and color depth & JavaScript &
     1920x1200x24\\ \hline
      List of fonts & Flash or JS &
     Abyssinica SIL,Aharoni CLM,AR PL UMing CN,AR PL UMing HK,AR PL UMing TW...\\ \hline
     List of HTTP headers & HTTP headers &
     Referer X-Forwarded-For Connection Accept Cookie Accept-Language 
     Accept-Encoding User-Agent Host \\ \hline
     Platform & JavaScript & Linux x86\_64 \\ \hline
	 Do Not Track & JavaScript & yes \\ \hline
     Canvas & JavaScript & \vspace{0.2cm}
     \includegraphics[width=8cm]{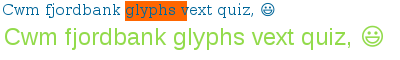}\\ \hline
     WebGL Vendor & JavaScript &
     NVIDIA Corporation\\ \hline
     WebGL Renderer & JavaScript &
     GeForce GTX 650 Ti/PCIe/SSE2\\ \hline
	 Use of an ad blocker & JavaScript & yes\\ \hline
			 
\end{tabular}
}

\end{table*}

\subsubsection{Adding new attributes}
Coupled with the seminal study from Eckersley, the enrichment of browser features prompted many different works aimed at adding new information in browser fingerprints and tracking them over time.

\paragraph{Canvas}
The Canvas API ``provides objects, methods, and properties to draw and manipulate graphics on a canvas drawing surface''~\cite{apiCanvas}.
Users can draw and animate any number of shapes and they can render textual content directly in the browser by using the graphical capabilities of the device. 
In 2012, Mowery and Shacham were the first to study the canvas API and the canvas 2D context in their Pixel Perfect study~\cite{mowery12PixelPerfect} to produce fingerprints.
As the font handling stacks can vary between devices, they state that the operating system, browser version, graphics card, installed fonts, sub-pixel hinting, and antialiasing all play a part in generating the final user-visible bitmap.
From 300 canvas samples using a specially crafted test, they observed 50 distinct renderings with the largest cluster containing 172 samples.

Later, Acar et al. performed a large-scale study of canvas fingerprinting in ``The Web Never Forgets''~\cite{acar14neverForgets}.
They found that scripts utilize the techniques outlined by Mowery and Shacham and notably, they take advantage of the fallback font mechanism of modern browsers to generate even more differences between devices.
This is the first time that such techniques were reported on the Internet.
They also noticed that most scripts share a very similar codebase and they explained this similarity by the availability on GitHub of an open source fingerprinting library called \textit{fingerprintjs}~\cite{fingerprintjs}.

\paragraph{WebGL}
The Khronos Group~\cite{apiWebGL} introduced WebGL which is a graphics API that can render interactive 3D objects in the browser and manipulate them through JavaScript without the need for plugins. 
Mowery and Shacham also studied in Pixel Perfect~\cite{mowery12PixelPerfect} the use of WebGL for fingerprinting.
In their test, they created a 3D surface on which they apply a very specific image and they add different ambient lights.
They observed 50 distinct renders from 270 samples.
They explain this heterogeneity by the difference in hardware and software where the processing pipeline is not exactly identical between devices.
However, it is not until 2017 that progress was made with regards to the capabilities of WebGL for fingerprinting.

Cao et al. designed a fingerprinting technique that relies heavily on WebGL to identify devices~\cite{cao17cross}.
Through a series of 31 rendering tasks, they test carefully selected computer graphics parameters to extract device features and they were able to uniquely identify more than 99\% of 1,903 tested devices.

\paragraph{AudioContext}
The Web Audio API provides an interface to create a pipeline to process audio~\cite{apiAudio}. 
By linking audio modules together, anyone can generate audio signals and apply very specific operations like compression or filtering to generate a very specific output.
Discovered by Englehardt et al. while crawling the web looking for trackers, AudioContext fingerprinting is one of the latest additions in a fingerprinter's toolbox~\cite{englehardt2016OpenWPM}.
They found scripts that process an audio signal generated with an \textit{OscillatorNode} to fingerprint devices.
The authors add that the fingerprinting process is similar to what is done with canvas fingerprinting as processed signals will present differences due to the software and hardware stack of the device.
The relative novelty of this technique explains that scripts using this API were only found on a very small number of websites.

\paragraph{Browser extensions}
Modern browsers embraced user customization from the beginning by allowing anyone to develop their own browser extensions, small add-ons that can extend the functionalities of a browser.
The most popular ones today are ad blockers, password managers or video downloaders~\cite{firefoxExtension}.
Detecting a browser extension is challenging as there is no API to query to get the exact list of installed extensions in the browser.
However, because of the way addons are integrated in browsers, it is possible to detect some of them.
A study conducted by Sj\"{o}sten et al. looked at the use of web accessible resources to detect extensions~\cite{sjosten2017Extension}.
By accessing very specific URLs, they can know if an extension is installed or not.
For example, to display the logo of an extension, the browser knows where it is stored on the device and it follows a URL of the form ``extension://\textless extensionID\textgreater/\textless pathToFile\textgreater'' to fetch it.
However, since these resources can be accessed in the context of any web page, this mechanism can be abused by a script to detect the presence or absence of a particular extension.
Not every extension has such accessible resources, and thus not every extension is detectable with this technique. 
Sj\"{o}sten et al. were able to detect 12,154 Chrome extensions out of 43,429 and 1,003 Firefox ones out of 14,896.

A second study was done by Starov and Nikiforakis and consists in identifying side effects produced by extensions~\cite{starov17extension}.
For example, if an extension adds a button on YouTube to provide new controls over a video, the added button is detectable by analyzing the DOM of the web page (the Document Object Model represents the structure of a page).
Detecting an ad blocker is similar as the blocking of an ad script will prevent some ads from being displayed.
The authors of the study performed an analysis of the 10,000 most popular Chrome extensions.
They found that 9\% of them produce DOM modifications that can be detected on any domain and 16.6\% introduce detectable changes on popular domains. In the user study based on 854 users and by detecting 1,656 extensions, Starov and Nikiforakis concluded that 14.10\% of users are unique.

A third study by S{\'{a}}nchez{-}Rola et al.~\cite{Sanc-etal-17-USENIX} used a timing side channel attack to detect browser extensions: they query resources of fake and existing extensions and measure the time difference between the calls. 
By using this method, they claimed to be able to detect any browser extension. S{\'{a}}nchez{-}Rola et al. have also conducted a user study of 204 users and aimed at detecting 2,000 extensions: they found that 56.86\% of users are unique.


Recently, Guly\'as et al.~\cite{Guly-etal-18-WPES} have conducted the biggest user study with 16,393 participants to evaluate uniqueness of users based on their browser extensions and Web logins. They used the method of Sj\"{o}sten et al.~\cite{sjosten2017Extension} based on web accessible resources and evaluated that out of 7,643 Chrome users, 39.29\% of them are unique based on the detection of 16,743 Chrome extensions. Additionally, Guly\'as et al. found out that it is sufficient to test only 485 carefully chosen extensions in order to achieve the same level of uniqueness.

Finally, Starov et al.~\cite{starov19bloat} looked at extension's bloat, i.e. the unnecessary side-effects caused by faulty application logic that reveal an extension's presence without providing any useful functionality.
These include injecting empty placeholders, injecting script or style tags, or sending messages on a page.
Out of 58,034 extensions from the Google Chromestore, they found that 5.7\% of them contained fingerprintable bloat and 61\% of them can be uniquely identified because of their bloat.
They propose a client-side access control mechanism for Chrome to protect users against bloat by controlling the reach of extensions.

\paragraph{JavaScript standards conformance}
Muzanni et al. proposed a method to reliably identify a browser based on the underlying JavaScript engine~\cite{mulazzani2013fast}. 
They analyzed browsers to see if they complied with the JavaScript standard and they tested them to detect which features were supported.
By collecting a dataset from more than 150 browser and operating system combinations, they were able to compute the minimal suite of tests that needs to be run to uniquely identify each combination.
Their approach is possible because web browsers present differences in the JavaScript engine even between two subsequent versions.

At the same time, Nikiforakis et al. explored the same idea by analyzing the mutability of the \textit{navigator} and \textit{screen} objects~\cite{nikiforakis13Cookieless}.
They highlighted that they can not only distinguish between browser families and versions but even between minor versions of the same browser.

Finally, much later, Schwarz et al. extended on this idea to go beyond the browser and find information on the system itself~\cite{schwarz19template}. 
They enumerated as many properties as they could in an automated fashion to find information that could reveal a difference on the OS and architecture levels .

\paragraph{CSS querying}
Unger et al. performed a series of test to detect CSS properties that are unique to some browsers~\cite{unger2013shpf}. 
For example, Firefox presents CSS properties prefixed with ``-moz-''~\cite{cssPrefixMozilla} while Chrome and Safari have some properties prefixed with ``-webkit-''~\cite{cssPrefixW3C}.
With their method, they can easily detect the browser family as these prefixes are not shared between browsers.

Saito et al. used the same technique to identify the browser and its family but they also went further by collecting information about the screen and the installed fonts through the use of \textit{@media} and \textit{@font-face} queries~\cite{cssSheets}.

\paragraph{Font metrics}
Fifield et al. looked into the analysis of character glyphs to identify devices on the web~\cite{fifield2015fontMetrics}.
They noticed that the same character with the exact same style may be rendered with different bounding boxes depending on the browser and the device used.
By testing 125,000 different Unicode characters on more than 1,000 web browsers, they were able to uniquely identify 34\% of their population.
With the data they collected, they were able to reduce the number of tested characters to 43 to reach the same conclusion.

\paragraph{Benchmarking} 
Another way to uncover information about a device is to benchmark its CPU and GPU capabilities.
Through JavaScript, a script can launch a series of tasks and measures the time it takes to complete them.
However, the biggest difficulty when using benchmarks is to interpret differences and fluctuations correctly.
Two time values can be different because they have been collected from two different devices but they could also belong to a single device where a new background process came disrupting the actual measurements.

Mowery et al. used 39 different tests to identify the performance signature of the browser's JavaScript engine~\cite{mowery11jsFP}.
They showed that they are able to detect the browser and its version with a 79.8\% accuracy.
However, the biggest downside of their approach is that it takes in total 190.8 seconds to run the complete benchmark suite. 
Contrarily to the majority of the attributes presented in this section that can be collected in a matter of milliseconds, this time difference makes it almost impossible to deploy such methods in the wild. 

Nakibly et al. turned to the WebGL API to display complex 3D scenes and measure the number of frames rendered by the browser~\cite{nakibly15hardware}.
They showed that benchmarking the GPU can produce very noticeable differences between devices as a small GPU on a smartphone will behave very differently than the latest high-end graphic card.

Finally, Saito et al. tried estimating the presence of specific CPU features like AES-NI and TurboBoost through benchmarking~\cite{cpuFeatures}.
Saito et al. went even further by identifying the CPU family and the number of cores with high accuracy~\cite{cpuFeatures2}.

Finally, S{\'{a}}nchez{-}Rola et al. measured clock difference on a device to perform device-based fingerprinting~\cite{SanchezRola18Clock}.
With native code, the authors can differentiate devices with the same hardware and software by measuring the time it takes to execute certain functions like ``string::compare'', ``std::regex'' or ``std::hash''. 
For the web implementation, they rely on the ``Crypto.getRandomValues()'' function and while they cannot differentiate all devices, it offers better results than Canvas or WebGL fingerprinting. 
However, the main problem here is that it is unclear if their approach can still work today with recent changes to the JavaScript Performance API because of side-channel attacks as no information is provided on the browser versions being used in the paper. 
There is also no information on the impact of the CPU load or the OS on the returned result.

\paragraph{Battery Status}
Drafted as early as 2011~\cite{batteryDraft}, the ``Battery Status'' specification defines ``an API that provides information about the battery status of the hosting device''~\cite{batterySpecification}. 
The API is composed of a \textit{BatteryManager} interface that reports if the device is charging or not. 
It also includes extra information like the charge level of the device along with its remaining charging and discharging time.
As detailed by the W3C, giving knowledge of the battery status to web developers would lead to power-efficient applications.
The intent behind the addition of this API seemed entirely legitimate.

However, they underestimated how much information regarding the battery could be misused in the wild.
In 2015, Olejnik et al. performed a privacy analysis of the Battery Status API~\cite{olejnik15battery}.
They highlighted the fact that the level of the battery could be used as a short-term identifier across websites and that repeated readouts could help determine the capacity of the battery.
The persons responsible for the standard did not anticipate all these problems as they only indicated in their original draft that the
``the information disclosed has minimal impact on privacy or fingerprinting''~\cite{batteryRecommendation}.
In order to address the issues posed by this API, many browser vendors decided to either remove this API~\cite{mozillaBatteryRemoval,webkitBattery} or spoof the given information~\cite{yandexBattery}.
Olejnik et al. documented extensively the complete history of the Battery Status API in~\cite{olejnik17privacy}.

\paragraph{Evolution of browser fingerprints over time}
Another core aspect of browser fingerprints concerns their evolution over time.
As a fingerprint is a direct reflection of a user's device and its environment, it is highly prone to changes as system's components are modified, configured or updated.
In order to enable long-term tracking, one must have the ability to understand these changes and anticipate how a fingerprint can change.

Eckersley was the first to look at this question in his Panopticlick study~\cite{eckersley2010Pano}.
He implemented an algorithm to heuristically estimate whether a given fingerprint might be an evolved version of a fingerprint seen previously.
From the fingerprints he collected, he was able to make a correct guess in 65\% of cases but he noted that the algorithm was ``very crude''.

Vastel et al. recently performed a more extensive study with FP-Stalker~\cite{vastel18stalker}.
Thanks to browser extensions for Firefox and Chrome, they collected fingerprints daily from volunteers and were able to witness first-hand the different changes a browser fingerprint can go through.
They first identified three different types of evolution: \textit{automatic evolutions} caused by organic software updates, \textit{context-dependent evolutions} reflected by changes in the user's environment, and \textit{user-triggered evolutions} caused by a change in the browser preferences.
They observed that the evolution of a fingerprint strongly depends on the device's type and how it is utilized.
At least one change was observed for 45.52\% of the collected fingerprints after one day, while it can take several weeks for other devices to see a single change.
Finally, they tried linking fingerprints belonging to the same device over time.
By collecting a fingerprint every three days, the algorithm behind FP-Stalker was capable of tracking a device for 51.8 days on average.
They were also able to track 26\% of devices for more than 100 days, proving that browser fingerprinting can effectively be used to complement other methods of identification.

\subsection{Analysing uniqueness of fingerprints}
\label{subsec:surveyors}

One of the most important aspects of browser fingerprints is their uniqueness.
If one device presents a combination of values that is unique, the impact on its privacy is strong as no stateful identifiers are required to track its whereabouts online. 
Here, we look at the different measurements used in the literature to asses the effectiveness of fingerprinting techniques.

\subsubsection{Evaluating fingerprinting techniques}

\paragraph{Entropy}
Entropy is used to quantify the level of identifying information in a fingerprint. 
The higher the entropy is, the more unique and identifiable a fingerprint will be.

Let $H$ be the entropy, $X$ a discrete random variable with possible values $\{x_1,...,x_n\}$ and $P(X)$ a probability mass function.
The entropy follows this formula:
$$H(X) = -\sum_{i} P(x_i) \log_bP(x_i) $$
In fingerprinting studies, the entropy of Shannon is in bits where $b=2$. 
One bit of entropy reduces by half the probability of an event occurring.

\paragraph{Normalized Shannon's entropy}
To compare datasets which are of different sizes, a normalized version of Shannon's entropy is used:
$$\frac{H(X)}{H_M}$$
$H_M$ represents the worst case scenario where the entropy is maximum and all values of an attribute are unique ($H_M = \log_2(N)$ with $N$ being the number of fingerprints in our dataset).
The advantage of this measure is that it does not depend on the size of the anonymity set but on the distribution of probabilities.
The quality of our dataset is quantified with respect to an attribute's uniqueness independently from the number of fingerprints in a database. 
This way, datasets can be compared despite their different sizes.

\paragraph{Anonymity sets}
One way adopted in the literature to visualize the distribution of collected fingerprints is through anonymity sets.
They give a direct representation of the distribution of a dataset by showing how devices or attributes with identical values are clustered together.
Notably, they can be used to show if one attribute has many different values spread evenly across different devices or if the majority of devices share a single unique value.
It can also be used to quantify how much protection can be provided by a defense mechanism by analyzing the difference in sets with and without the protection.

\subsubsection{Large scale studies}
We detail here the only three large scale studies performed on the effectiveness of tracking with browser fingerprinting.
Table~\ref{table:overview} provides an overview of three large scale fingerprinting studies, while Table~\ref{table:entropy} gives a summary of the attributes along with their detailed entropy numbers.

\begin{table}[h!]
\center
\caption{Overview of large scale studies on browser fingerprinting.}
\label{table:overview}

\begin{tabular}{c|M{2cm}|*{2}{M{1.2cm}| } M{1.2cm} | M{1.2cm} }
\multirow{2}{*}{}                            & Panopticlick (2010) & \multicolumn{2}{M{2.4cm}|}{AmIUnique (2016)} & \multicolumn{2}{M{2.4cm}}{Hiding in the Crowd (2018)} \\ \cline{2-6} 
                                             & Desktop             & Desktop              & Mobile             & Desktop             & Mobile           \\ \hline
\multicolumn{1}{M{2cm}|}{Number of fingerprints} & 470,161             & 105,829            & 13,105           & 1,816,776         & 251,166        \\ \hline
\multicolumn{1}{M{2cm}|}{Unique fingerprints}    & 94.2\%              & 89.4\%             & 81\%             & 35.7\%            & 18.5\%         \\
\end{tabular}

\end{table}

\paragraph{Panopticlick}
As detailed in Section~\ref{subsec:pioneers}, Eckersley conducted the Panopticlick experiment in 2010~\cite{eckersley2010Pano}.
From 470,161 fingerprints, he concluded that browser fingerprinting can be used to track users online as 83.6\% of collected fingerprints were unique.
This number rose to 94.2\% if users had enabled Flash or Java.
The most discriminating attributes at the time were the list of plugins, the list of fonts and the user-agent.

\paragraph{AmIUnique}
Laperdrix et al. performed an analysis of 118,934 fingerprints in 2016~\cite{laperdrix16Beauty} and their study brought to light new results.
First, they confirm Eckersley's findings from 2010 as 89.4\% of their collected fingerprints were unique.
However, in the 6 years that separated both studies, they saw an evolution in the different attributes that compose a fingerprint.
While the list of plugins and fonts were kings at the beginning of the decade, it is not the case anymore as plugins have been deprecated in major browsers because of the security threat they pose~\cite{npapi,pluginEndFirefox}.
Newcomers like canvas fingerprinting provide very strong results as they observed an important entropy in the collected values.
Then, at a time where the use of smartphones is booming, they show that mobile fingerprinting is possible but for different reasons than on desktops.
In their dataset, 81\% of fingerprints from mobile devices are unique.
HTTP headers and HTML5 canvas fingerprinting play an essential role in identifying browsers on these devices.
Finally, they simulate scenarios to assess the impact of future web evolutions. 
They show that certain scenarios would limit the detriment these technologies have on privacy, while preserving the current trend towards an ever more dynamic and rich web. 
In their study, simple changes like having generic HTTP headers or removing plugins reduce fingerprint uniqueness in desktop fingerprints by a strong 36\%.

\paragraph{Hiding in the Crowd}
In 2018, G\'omez-Boix et al. analyzed 2,067,942 fingerprints collected on one of the top 15 French websites~\cite{gomez18hiding}.
Their findings provide a new layer of understanding to the domain as 33.6\% of fingerprints from their dataset were unique.
Compared to the other two large scale studies, this number is two to three times lower.
When considering mobile devices, the difference is even larger as 18.5\% of mobile fingerprints were unique compared to the 81\% from~\cite{laperdrix16Beauty}.
Their study highlights the importance of the data collection process.
In the past, fingerprints have been collected on websites that explicitly target visitors who are aware of online privacy or who might be more cautious than the average web user.
Here, their data is collected on a commercial website targeting a more global audience.
This characteristic of the dataset coupled with the very high number of collected fingerprints are the keys to understand the differences in fingerprint uniqueness.
They also demonstrate that desktop fingerprints are mostly unique because of their combination of attributes whereas mobile devices present attributes that have unique values.


%
\begin{table*}[h!]
\center
\caption{Browser attributes, their entropy and their normalized entropy from the Panopticlick~\cite{eckersley2010Pano}, AmIUnique~\cite{laperdrix16Beauty} and Hiding in the Crowd~\cite{gomez18hiding} studies.}
\label{table:entropy}

\resizebox{\textwidth}{!}{%
\begin{tabular}{ M{3cm} |*{5}{M{1.6cm}|} M{1.6cm} }
\multirow{2}{*}{Attribute} & \multicolumn{2}{c|}{Panopticlick (2010)}& \multicolumn{2}{c|}{AmIUnique (2016)} & \multicolumn{2}{c}{Hiding (2018)} \\ \cline{2-7}
& Entropy & Normalized entropy & Entropy & Normalized entropy & Entropy & Normalized entropy \\ \hline

	User agent & 10.000 & 0.531 & 9.779 & 0.580 & 7.150 & 0.341\\ \hline
     Accept & - & - & 1.383 & 0.082 & 0.729 & 0.035\\ \hline
      Content encoding & - & - & 1.534 & 0.091 & 0.382 & 0.018 \\ \hline
      Content language & - & - & 5.918 & 0.351 & 2.716 & 0.129 \\ \hline
      List of plugins & 15.400 & 0.817 & 11.060 & 0.656 & 9.485 & 0.452\\  \hline
      Cookies enabled & 0.353 & 0.019 & 0.253 & 0.015 & 0.000 & 0.000\\ \hline
      Use of local/session storage & - & - & 0.405 & 0.024 & 0.043 & 0.002\\ \hline
      Timezone & 3.040 & 0.161 & 3.338 & 0.198 & 0.164 & 0.008\\ \hline
      Screen resolution and color depth & 4.830 & 0.256 & 4.889 & 0.290 & 4.847 & 0.231\\ \hline
      List of fonts & 13.900 & 0.738 & 8.379 & 0.497 & 6.904 & 0.329\\ \hline
     List of HTTP headers & - & - & 4.198 & 0.249 & 1.783 & 0.085\\ \hline
     Platform & - & - & 2.310 & 0.137 & 1.200 & 0.057\\ \hline
	 Do Not Track & - & - & 0.944 & 0.056 & 1.919 & 0.091\\ \hline
     Canvas & - & - & 8.278 & 0.491 & 8.546 & 0.407\\ \hline
     WebGL Vendor & - & - & 2.141 & 0.127 & 2.282 & 0.109\\ \hline
     WebGL Renderer & - & - & 3.406 & 0.202 & 5.541 & 0.264\\ \hline
	 Use of an ad blocker & - & - & 0.995 & 0.059 & 0.045 & 0.002\\ \hline
			 
	$H_M$ (worst scenario) & \multicolumn{2}{c}{18.843} & \multicolumn{2}{|c}{16.860} & \multicolumn{2}{|c}{20.980} \\\hline
	Number of FPs & \multicolumn{2}{c}{470,161} & \multicolumn{2}{|c}{118,934} & \multicolumn{2}{|c}{2,067,942} \\ 

\end{tabular}
}

\end{table*}

\subsection{Adoption of fingerprinting on the web}
\label{subsec:watchers}

Since Eckersley's study in 2010, different studies have been conducted to quantify the adoption rate of browser fingerprinting on the web. 
Table~\ref{tab:adoption} provides an overview of the major four studies.

\begin{table*}[h!]
\center
\caption{Overview of four studies measuring adoption of browser fingerprinting on the web.}
\label{tab:adoption}
\resizebox{\textwidth}{!}{%
\begin{tabular}{ M{2cm} | M{3cm} | M{3cm} | M{3cm} | M{5cm} }
 & Fingerprinting techniques detected & Sites crawled & Prevalence & Detection method \\ \hline
Cookieless Monster \cite{nikiforakis13Cookieless} (2013) & Detection of 3 known fingerprinting libraries & 10K sites (up to 20 pages per site) & 0.4\% & Presence of JS libraries provided by BlueCava, Iovation and ThreatMetrix. \\ \hline
FPDetective \cite{acar13fpDet} (2013) & JS-based and Flash-based font probing & 1M sites (homepages)\newline 
100K sites (25 links per site) for JS \newline
10K (homepages) for Flash & 0.04\% (404 of 1M) for JS-based \newline 1.45\% (145 of 10K) for Flash-based & 
Logging calls of font probing methods. 
A script that loads more than 30 fonts or a Flash file that contains font enumeration calls is considered to perform fingerprinting.\\ \hline
The Web Never Forgets \cite{acar14neverForgets} (2014) & Canvas fingerprinting & 100K sites (homepages) & 5.5\% & Logging calls of canvas fingerprinting related methods. 
A script is considered to perform fingerprinting 
if it also checks other FP-related properties.\\ \hline
1M Alexa study with OpenWPM \cite{englehardt2016OpenWPM} (2016) & Canvas fingerprinting, canvas-based font probing, WebRTC and AudioContext & 1M sites (homepages) & 1.4\% for canvas fingerprinting \newline 0.325\% for canvas font probing \newline 0.0715\% for WebRTC \newline 0.0067\% for AudioContext & Logging calls of advanced FP-related JavaScript functions.\\ \hline
10K Majestic study \cite{alFannah18amensia} (2018) & 17 attributes (including OS, screen, geolocation, IP address among others) & 10K sites (homepages) & 68.8\% & 
Data leaving the browser must contain at least one of the 17 monitored attributes.

\end{tabular}
}

\end{table*}

In 2013, Nikiforakis et al. with the Cookieless Monster study~\cite{nikiforakis13Cookieless} crawled up to 20 pages 
for each of the the Alexa top 10,000 sites to look for fingerprinting scripts from the three following companies: 
BlueCava, Iovation, ThreatMetrix.
They discovered 40 sites (0.4\%) making use of these companies' 
fingerprinting code.

The same year, Acar et al. performed a much larger crawl by visiting the homepages of top Alexa 1 million websites 
and 25 links of 100,000 Alexa websites with the FPDetective framework~\cite{acar13fpDet}.
They made modifications to the rendering engine to intercept and log access to browser and 
device properties that could be used for fingerprinting. 
They decompile the Flash files they encounter during crawling to verify the presence of 
fingerprinting related function calls.
FPDetective study was the first to measure adoption of fingerprinting scripts without 
relying on a known list of tracking scripts as they directly looked for behaviors related to fingerprinting activities.
They found 404 sites out of 1 million performing JavaScript-based font probing and 145 sites out of 10,000 performing Flash-based font probing.

In 2014, Acar et al. performed the ``The Web Never Forgets'' study~\cite{acar14neverForgets}, 
where they measured adoption of canvas fingerprinting on homepages of 100,000 Alexa websites. 
They instrumented the browser 
to intercept calls and returns to Canvas related methods, 
and tried to remove false positives   
by a set of rules (more details in Section 3.1 of~\cite{acar14neverForgets}).  
They found 5542 sites out of 100,000 performing Canvas fingerprinting. 

In 2016,  
Engelhardt and Narayanan released the OpenWPM platform, ``a web privacy measurement framework 
which makes it easy to collect data for privacy studies on a scale of thousands to millions of websites''~\cite{repoOpenWPM}. 
To demonstrate the capabilities of their tool, they made an analysis of the Alexa top 1 million sites to 
detect and quantify emerging online tracking behaviours~\cite{englehardt2016OpenWPM}. 
Their findings provide more accurate results than in the past as they instrumented extensively a very 
high number of JavaScript objects to build a detection criterion for each known fingerprint technique 
(more details in Section 3.2 of~\cite{englehardt2016OpenWPM}).
Out of 1 million websites, they found 14,371 sites performing canvas fingerprinting, 
3,250 sites performing canvas font fingerprinting, 
715 sites performing WebRTC-based fingerprinting, 
and only 67 sites performing AudioContext fingerprinting. 

Finally, in 2018, Al-Fannah et al. crawled the top Majestic 10,000 websites and recorded what was sent out by the browser~\cite{alFannah18amensia}.
Their definition of fingerprinting is much broader and inclusive than the other studies presented in this section.
A website is deemed to be performing fingerprinting if at least one attribute out of a list of 17 is present in the recorded payloads.
They identified 6,876 (68.8\%) websites as performing fingerprinting which is a much higher number than what was reported in the past. 
84.5\% of them are third parties and, by analyzing what was collected, the authors identified in total 284 attributes that can be used for fingerprinting.

\paragraph{Challenges in measuring adoption}

In order to quantify the number of websites that are currently using fingerprinting scripts on the Internet, one needs the means to identify them.
However, even if the collection process in a browser is straightforward, the reality is in fact much more complex.
If a script accesses the user-agent header and the list of plugins, it could be for legitimate purposes to tailor the current web page to the user's device.
But it could also be the first-step towards building a complete browser fingerprint.
If a script makes hundreds of calls to the Canvas API, it may be rendering a 2D animation in the browser. 
But it may also probe for the list of fonts installed on the system.
These simple examples illustrate that the line between a benign script and a fingerprinting one is far from being clearly defined.
As we saw in the previous paragraph, the protocol to identify a fingerprinting script can lead to very different numbers on the adoption of this technique.
Researchers are currently facing a lot of challenges to classify scripts correctly as the goal of two scripts can vastly vary even if they present very similar content.

Here, we list some telltale signs that indicate that a script may be partaking in fingerprinting activities.

\begin{itemize}

\item \textbf{Accessing specific functions}
In the fingerprinting literature, many functions and objects are known to return device-specific information (see Table~\ref{table:example}).
For example, the \textit{navigator} object contains the user-agent and the platform.
Does the script access these very specific functions and objects? 

\item \textbf{Collecting a large quantity of device-specific information}
Even if a script access the screen resolution, this information alone is not sufficient to identify a device on the Internet.
If a script queries specific APIs, how many of them are accessed?
Can the collected information be used to identify a single device?
Studies like~\cite{acar14neverForgets,englehardt2016OpenWPM} have looked specifically at APIs like Canvas, WebRTC or AudioContext.
They did not consider the full list of attributes that could be collected to assess if a script is performing fingerprinting or not.

\item \textbf{Performing numerous access to the same object or value}
If a function is called an incredible number of times, can it be considered as a normal usage of the API? 
Or is the script testing different parameters to expose a certain property of the device?
How can we consider a usage as normal or abnormal?

\item \textbf{Storing values in a single object}
Is the script storing all collected values in the same object?
From a design perspective, having all the values in the same object means that they probably share a similar purpose.

\item \textbf{Hashing values}
Scripts can hash very long strings to ease processing, transfer or server-side storage. 
The popular\textit{fingerprintjs} library~\cite{fingerprintjs} as a default option hashes the entirety of the device's fingerprint.
Is the script hashing any value, especially ones that come from known fingerprinting functions?

\item \textbf{Creating an ID}
Does the script generate a string that looks like an identifier?
Is this ID stored in a cookie or in any cache mechanisms of the browser?

\item \textbf{Sending information to a remote address}
Are there any pieces of data containing device-specific information sent to a remote server?

\item \textbf{Minification and Obfuscation}
``Minifying'' a script consists in removing all unnecessary characters from its source code like white space characters, new line characters or comments without changing its functionality.
A lot of well-known JavaScript libraries are ``minified'' to reduce the amount of data that needs to be transferred when they are downloaded.
For example, the weight of the famous jQuery library~\cite{jquery} in version 3.3.1 is cut in three just by minifying the code (from 271.8kb to 86.9kb).
Figure~\ref{lst:fibonacci} shows a simple implementation of the Fibonacci sequence in JavaScript.
The minified version is much more compact than the original version.

On top of minification, a JavaScript file can be obfuscated, \ie modified to make it difficult to read and understand.
Some variables can be renamed to very short and meaningless names.
Some sections can be intertwined to make it difficult to follow the flow of the program.
Some parts of the code can also self-generate the true payload similar to what is observed with packing mechanisms in malwares.
Most developers use obfuscation to protect their source code and to prevent other developers from copying it but others see it as a way to hide the true meaning of their code.
In the end, it requires reverse-engineering efforts to know the true intent of the author and it requires far more means to correctly find if a script is conducting fingerprinting activities.

\begin{figure}[h!]
\begin{minipage}{\columnwidth}
\centering
\begin{lstlisting}[language=JavaScript,title={Standard},label={lst:min1}]
function fib(n) {
    if(n <= 1) {
        return n;
    } else {
        return fib(n - 1) + fib(n - 2);
    }
}
\end{lstlisting}
\end{minipage}
\hfill
\begin{minipage}{\columnwidth}
\centering
\begin{lstlisting}[title={Minified},label={lst:min2}]
function fib(a){return a<=1?a:fib(a-1)+fib(a-2)}
\end{lstlisting}
\end{minipage}

\begin{minipage}{\columnwidth}
\centering
\begin{lstlisting}[title={Obfuscated by~\cite{jsObfuscator}},label={lst:min3}]
eval(function(p,a,c,k,e,d){e=function(c){return c};if(!''.replace(/^/,String)){while(c--){d[c]=k[c]||c}k=[function(e){return d[e]}];e=function(){return'\\w+'};c=1};while(c--){if(k[c]){p=p.replace(new RegExp('\\b'+e(c)+'\\b','g'),k[c])}}return p}('4 3(0){5 0<=1?0:3(0-1)+3(0-2)}',6,6,'a|||fib|function|return'.split('|'),0,{}))
\end{lstlisting}
\end{minipage}

\caption{JavaScript code for the Fibonacci sequence. The three pieces of code are all equivalent.}
\label{lst:fibonacci}
\end{figure}

\end{itemize}

In the end, fine tuning all of these rules and identifying a script as a fingerprinting one present many difficulties.
Engelhardt and Narayanan noted in~\cite{englehardt2016OpenWPM} that a large number of fingerprinting scripts were not blocked by popular privacy tools, especially the lesser known ones.
The number of actors actually performing device fingerprinting on the web may very well be much higher than what is currently reported by large crawls.
\section{Defense techniques}
\label{sec:defense}

In this section, we detail techniques and solutions aimed at mitigating the effects of browser fingerprinting.
The goal is to improve users' privacy by preventing unwanted tracking.
As we will see, there is no ultimate approach that can prevent fingerprinting while keeping the richness of a modern web browser.
Designing a strong defense requires a fine-tuned balance between privacy and usability that can be challenging to get right.

Table~\ref{background:table:defenses} provides a summary of all the defenses detailed in this section.
While some solutions provide very strong protection against browser fingerprinting, it is often at the cost of usability as we can see for example with NoScript or the Tor Browser.
From the scientific publications, we can see that the biggest challenge met by researchers is to provide a complete coverage of modified attributes as the slightest mismatch render users more visible to trackers.
A solution can be rendered useless in a matter of months as browsers are constantly updated and new APIs are surfacing frequently.

\newcommand{\comments}{10.0cm}
\newcommand{\plusminus}[2]{\parbox[c]{\comments}{\vspace{.3\baselineskip} $+$: #1\\$-$: #2 \vspace{.3\baselineskip}} }

\begin{table}[ht!]
\center
\caption{Summary of existing defense solutions.
M = Modifying the fingerprint content. 
M* = Modifying the fingerprint content by switching browsers.
U = Universal fingerprint.
BS = Blocking Scripts.
BA = Blocking APIs or Access.
}
\label{background:table:defenses}
\scalebox{0.78}{
\begin{tabular}{M{0.5cm} | M{2.5cm} | M{1.5cm} | M{0.9cm} | M{\comments} }
     \multicolumn{1}{c|}{\multirow{1}{*}{}} & Solution & Ref. & Type & Comments\\\hline
     \multirow{20}{*}{\rotatebox[origin=c]{90}{Scientific publications}} & FP-Block & \cite{torres15FPBlock} & M 
	 & \plusminus{Separation of web identities}{Incomplete coverage} \\\cline{2-5}
     & FPGuard & \cite{faiz15FPguard} & M 
     & \plusminus{Detection and prevention of fingerprinting}{Lack of details} \\\cline{2-5}
     & Fiore et al. & \cite{fiore14countering} & M 
     & \plusminus{Aims at creating consistent fingerprints}{Incomplete coverage} \\\cline{2-5}
     & DCB & \cite{baumann16disguised} & M 
     & \plusminus{N:1/1:N strategies, changes at each session, creation of groups with similar configurations}{Incomplete coverage}  \\\cline{2-5}
     & PriVaricator & \cite{nikiforakis15Privari} & M 
     & \plusminus{Custom randomization policies}{Incomplete coverage} \\\cline{2-5}
     & Blink & \cite{laperdrix15Mitigating} & M 
     & \plusminus{Produces genuine and diverse fingerprints with no inconsistencies}{Takes HDD space} \\\cline{2-5}
	 & FPRandom & \cite{laperdrix17fprandom} & M 
	 & \plusminus{Introduces noise into the Canvas, AudioContext APIs and randomizes the enumeration order of JavaScript objects}{Other vectors can still be used} \\\cline{2-5}
     & Changing browsers & \cite{boda12crossBrowser,cao17cross} & M* 
     & \plusminus{Presents distinct and genuine fingerprints}{Can be bypassed} \\\cline{2-5}
     & Cliqz browser & \cite{yu16cliqz,cliqz} & BS 
     & \plusminus{Strong protection against scripts with unique identifiers}{Relies on a central server} \\\cline{2-5}
     & Latex Gloves & \cite{sjosten19extensions} & BA 
     & \plusminus{Protection against extension fingerprinting}{Relies on user-curated whitelists} \\\cline{2-5}     
     & CloakX & \cite{trickel19extensions} & M 
     & \plusminus{Strong protection against extension fingerprinting}{Other vectors an still be used} \\\cline{2-5}
     & UniGL & \cite{wu19webGL} & U
     & \plusminus{Protection against WebGL fingerprinting for all devices}{Other vectors can still be used} \\\cline{2-5} 
     \hline
          
	 \multirow{23}{*}{\rotatebox[origin=c]{90}{Online tools}} & Canvas Defender & \cite{canvasDefender} & M 
	 & \plusminus{Modifications consistent across a browsing session}{Only canvas} \\\cline{2-5}
     & Random Agent Spoofer & \cite{randomSpoofer} & M 
     & \plusminus{Uses real database of browser profiles}{Incomplete coverage} \\\cline{2-5}
     & Tor Browser & \cite{torBrowser} & U BA 
     & \plusminus{Very strong protection against fingerprinting }{Tor fingerprint is brittle} \\\cline{2-5}
     & NoScript & \cite{noscript} & BS 
     & \plusminus{Blocks all JavaScript scripts including fingerprinting scripts}{Blocks all JavaScript scripts including scripts needed to correctly display a webpage} \\\cline{2-5}
     & Filter list-based ad/tracker blocker & \cite{adblockplus,ghostery,ublockGithub,disconnect} & BS 
     & \plusminus{Extensive blocking list}{Relies on lists of known trackers} \\\cline{2-5}
     & Privacy Badger & \cite{privacyBadger} & BS 
     & \plusminus{Heuristics-based approach}{Blocking may be too aggressive} \\\cline{2-5}
	 & Canvas Blocker & \cite{canvasBlocker} & BA 
	 & \plusminus{Blocks the entire canvas API}{Other vectors can still be used} \\\cline{2-5}
	 & Brave browser & \cite{brave} & BA 
	 & \plusminus{Blocks the entire Canvas, WebGL, AudioContext and WebRTC APIs}{Other vectors can still be used} \\\cline{2-5}
	 & Firefox fingerprinting resistance & \cite{firefoxProtection,torUplift,firefoxFusion} & M BA U& \plusminus{Blocks the entire Canvas API and reduces the quantity of information on several attributes}{Still a work in progress, other vectors can still be used} \\
\end{tabular}
}
\end{table}

\subsection{Increasing device diversity}
\label{subsec:increasing}

\subsubsection{Modifying the content of fingerprints}
The first defense to mitigate browser fingerprinting is to increase the diversity of devices so that real fingerprints are hidden in noise.
The intuition behind this method is that third parties rely on fingerprint stability to link fingerprints to a single device.
By sending randomized or pre-defined values instead of the real ones, the collected fingerprints are so different and unstable that a tracker is unable to identify devices on the web.

\paragraph{The inconsistency problem}
While this approach can appear to be strong on paper, the reality is much more complicated as Peter Eckersley called it \textit{the Paradox of Fingerprintable Privacy Enhancing Technologies}~\cite{eckersley2010Pano}.
Instead of enhancing users' privacy, some tools make fingerprinting easier by rendering a fingerprint more distinctive.
By looking through the extensions available for both Chrome and Firefox, one can find many \textit{spoofers} or \textit{switchers} to modify the actual values that are collected by scripts.
One of the most popular ones on Firefox called Random Agent Spoofer~\cite{randomSpoofer} claims more than 100,000 users at the time of writing and it provides the ability to rotate ``complete browser profiles ( from real browsers / devices ) at a user defined time interval''.
Nikiforakis et al. performed an analysis of these extensions and found many issues with regards to browser fingerprinting~\cite{nikiforakis13Cookieless}.
These extensions can modify a property but may forget to change another one, creating a mismatch between attributes.
One browser could announce in its user-agent that the underlying OS is Linux while the \textit{navigator.platform} property indicates it is running on Windows.
Another example would be a device claiming to be an iPhone while the reported screen resolution is far bigger than what is currently supported on these devices.
While the idea of switching values with other ones is promising, the constant evolution of browsers coupled with very strong links between attributes prevent this approach from being recommended.
To fix the shortcomings of these agent spoofers, the scientific community turned itself to new approaches.

\paragraph{Replacing the values of attributes}
Torres et al. explored the concept of separation of web identities with a solution called FP-Block~\cite{torres15FPBlock}.
When the browser connects to a new domain, it will generate a new identity (\ie a new fingerprint) for this particular domain.
The intuition behind FP-Block is that third parties will see different fingerprints on each site they are embedded so that tracking is hampered.

FaizKhademi et al. developed the FPGuard solution which runs in two phases: detection and prevention~\cite{faiz15FPguard}.
First, they detect fingerprinting-related activities with a series of 9 metrics.
Then, from these metrics, they compute a suspicion score and if this score goes above a specific threshold, the second phase kicks in where a series of components will modify the content of the device fingerprint.

Fiore et al. worked to counter unwanted tracking by creating fingerprints that resemble the ones left by someone else~\cite{fiore14countering}. 
They claim that they have to alter data in a way that is consistent to prevent being detected.
They modify a very specific subset of fingerprintable attributes with a fake browsing profile.

Baumann et al. designed a solution to disguise the Chromium browser called DCB (Disguised Chromium Browser) by changing the following parameters: the screen resolution, the browser language, the user-agent, the time and date, the list of fonts and the list of plugins~\cite{baumann16disguised}.
When DCB launches, it contacts the main server that ``maintains a database of real world fingerprinting features to enforce a robust browser configuration on the client'' and then applies one of the two following strategies: ``N:1 Many Browsers, One Configuration'' or ``1:N One Browser, Many Configurations''.

Nikiforakis et al. explored with PriVaricator the use of randomization to render browser fingerprints unreliable for tracking~\cite{nikiforakis15Privari}.
Instead of blatantly lying as it can seriously degrade the user experience, they introduced the concept of randomization policies. 
Each policy details the modifications made to a specific attribute along with a set of requirements that define when it kicks in. 
This way, any developer can define her own modification strategy that balances effectiveness with usability. 

Trickel et al. designed CloakX, a solution based on client-side diversification to prevent the detection of installed extensions~\cite{trickel19extensions}.
The core idea behind CloakX is that it randomizes what makes an extension identifiable while maintaining equivalent functionality.
CloakX follows different steps to completely cloak an extension: it randomizes the path of web accessible resources to prevent probing attacks, it changes the behavioral fingerprint by changing ID and class names that are injected and it adds a proxy to perform the necessary mappings from dynamic calls.

Finally, Laperdrix et al. argued that the diversity of software components that is at the source of browser fingerprinting provides strong foundations for a counter measure~\cite{laperdrix15Mitigating}.
They proposed an original application of dynamic software reconfiguration techniques to establish a moving target defense against browser fingerprint tracking.
With a tool called Blink, they randomly assemble a coherent set of components (an OS, a browser, plugins, etc.) every time the user wants to browse the web.
Exposed fingerprints then break the stability needed for their exploitation as they are very different from each other.
The strongest advantage of Blink compared to other tools is that the exhibited fingerprints are genuine with no mismatches between attributes since they rely on real components running on the user's device.

\paragraph{Introducing noise}
While most attributes are collected in string form, other ones from the Canvas or AudioContext APIs produce more complex data structures.
Instead of simply replacing an output with another pre-defined one, one can introduce noise into the rendering process of these APIs.
This way, a Canvas or AudioContext test can be ever so slightly different at each execution.

One way to introduce noise is to position the modification at the very end of the processing pipeline where a script collect its values.
An extension called Canvas Defender on Firefox does exactly this~\cite{canvasDefender}.
When a script renders an image, the browser will behave normally and the user will see the intended image.
However, when the script tries to read the content of the rendered canvas element, it will go through a function that modifies the actual RGB values of each pixel.
The image collected by the script is then different from the image that the user can see.

Baumann et al. positioned themselves much earlier in the rendering pipeline by directly modifying the Chromium source code in DCB~\cite{baumann16disguised}.
They modified the \textit{fillText()} and \textit{strokeText()} that are heavily used in canvas fingerprinting scripts to alter the renderings of canvas elements at runtime.
Their approach also provides consistency in the same browsing session as they use a random session identifier generated at startup to steer the modifications.

Laperdrix et al. proposed with FPRandom to exploit browsers’ untapped flexibility to introduce randomness~\cite{laperdrix17fprandom}. 
Their goal was to increase non-determinism in browsers to reduce the side-effects that cause fingerprintable behaviours.
The authors modified the source code of Firefox to target modern fingerprinting techniques, namely canvas fingerprinting, AudioContext fingerprinting and the unmasking of browsers through the order of JavaScript properties.

\paragraph{The challenges in modifying the content of fingerprints}
This section showed that it is possible to increase the diversity of exposed fingerprints and modify their content but the challenges to have a working and undetectable solution are numerous.
Attributes cannot be modified in a way that will break browsing.
The slightest mismatch between two attributes can make a user more visible to trackers which defeats the entire purpose of running a defense solution.
All the techniques detailed here pose the question if such kind of approach should be explored further or if the constant evolution of web browsers render current implementations incredibly hard to maintain and to recommend.
Vastel et al. developed \textit{FP-Scanner}~\cite{vastel18scanner}, a test suite that explores browser fingerprint inconsistencies to detect potential alterations.
By applying a progressive detection logic from attributes collected at the browser level to the OS level, the scanner can detect if an attribute was modified and, to some extent, reveal the original unaltered value hidden by the countermeasure.
The authors also argue that detecting a fingerprinting countermeasure does not necessarily imply that a user can be tracked more easily as it depends on what information is being leaked by the defense mechanism and how stable it can be.
In the end, this article shows that, while researchers and developers are finding many ways to make fingerprints unstable, there are always really small details that are easy to overlook that make current solutions ineffective. 
Modern web browsers are such complex pieces of machinery that it is incredibly hard to predict where the next piece of revealing information will be.

\subsubsection{Changing browsers}
\label{subsubsec:cross}
Since a large part of a device fingerprint is composed of browser-specific information, one could decide to use two different browsers to have two distinct device fingerprints. 
This way, it is harder for a third party to have a complete picture of a user's browsing patterns as the tracking party will obtain two decorrelated browsing profiles.
While the premise behind this idea is really simple, the truth behind it is more complicated.
Two studies have shown that collecting attributes that are specific to the OS and the hardware can be sufficient to uniquely identify a device.

In 2012, Boda et al. were the first to design a browser-independent fingerprinting algorithm that rely mainly on attributes like the list of fonts, the timezone and the screen resolution~\cite{boda12crossBrowser}.
Their findings show that the list of fonts provide a solid base for identification and that they were able to identify returning visitors who used more than one browser or changed their IP addresses dynamically.
However, their dataset contained only 989 users, which was not diverse enough and therefore prevented them from concluding whether the same results hold at a larger scale.

In 2017, Cao et al. designed a fingerprinting technique that relies heavily on the OS and hardware functionalities of a device~\cite{cao17cross}.
By rendering 31 different tasks with the WebGL API, they are able to extract device features from carefully selected computer graphics tests and they show that they are able to uniquely identify devices even if the user switches browser.
One important detail is that their whole suite of tests take several seconds to be fully executed contrarily to more standard fingerprinting scripts which take less than a second.
However, their dataset was also relatively small (3,615 fingerprints from 1,903 users) and therefore more studies are needed to evaluate how unique users are based on this technique at a larger scale.

In the end, cross-browser fingerprinting is a reality even if its deployment in a real-world solution may prove very challenging mainly due to time constraints.
By collecting enough data from the OS and hardware layers of a system, a third party can uniquely identify a device.

\subsection{Presenting a homogeneous fingerprint}
\label{subsec:homogeneous}

Another defense strategy is to make all devices on the web present the same fingerprint.
This is the approach chosen by the Tor Browser~\cite{torBrowser} also known as TBB (the Tor Browser Bundle) which uses the Tor network.
Wu et al. also designed UniGL to remove the discrepancies between devices when performing WebGL fingerprinting.

\subsubsection{The Tor Browser}

\paragraph{The theory}
While the Tor network prevents an attacker from finding out the real IP address of a client, it does not modify the actual content of an HTTP request.
If a cookie ID or a browser fingerprint is present in the payload, a server can uncover the true identity of a user.
To fix this problem, the Tor Browser was developed.
As detailed by the official design document~\cite{torDesign}, the Tor Browser follows a set of requirements and one of them includes a \textit{Cross-Origin Fingerprinting Unlinkability} section which specifically targets browser fingerprinting.
While they acknowledge that randomization can be effective to protect against fingerprinting, they chose uniformity or the \textit{one fingerprint for all} strategy for Tor users.
The design document lists 24 different modifications that have been introduced in the Tor Browser.
The most notable ones are the blocking of the Canvas and WebGL API, the complete removal of plugins, the inclusion of a default bundle of fonts to prevent font enumeration and the modification of the user-agent along with HTTP headers.
Whether a user is on Windows, Mac and Linux, the Tor Browser will always report that the device is on Windows.

\paragraph{The reality}
While the Tor Browser can be considered as one of the strongest defenses against browser fingerprinting, it still presents some shortcomings.
The fingerprint exposed by the Tor Browser is known and easily identifiable.
Data like the user-agent, the screen resolution and the IP addresses from known Tor exit nodes are enough information to distinguish the Tor browser from a standard one.
While this may not be important with respect to identification as one website cannot distinguish one Tor user from another one, it can still impact their browsing experience as shown by Khattak et al.~\cite{khattak16differential}.
They reported that 3.67\% of the top 1,000 Alexa sites either block or offer degraded service to Tor users to reduce Internet abuse.

The second problem with Tor browser fingerprints is their brittleness as differences can still between browsers like the screen resolution.
When first launched, the Tor Browser window has a size of 1,000x1,000.
However, if the user decides to maximize the window, the browser displays the following message: ``Maximizing Tor Browser can allow websites to determine your monitor size, which can be used to track you. We recommend that you leave Tor Browser windows in their original default size.''.
If the user has an unusual screen resolution, this information could be used to identify her as she will be the only Tor user with this screen resolution.

The third problem is that detectable differences exist between operating systems running the Tor Browser.
The design document notes that they intend to reduce or eliminate OS type fingerprinting to the best extent possible but they add that the efforts in that area is not a priority. 
While this may provide very few information compared to other fingerprinting vectors, OS differences are yet an additional vector that can be used to distinguish a user from the pool of all Tor users.

In the end, developers of the Tor Browser have made some very strong modifications to limit the fingerprintability of the browser as much as possible.
If users stick with the default browser fingerprint that most users share, it provides the strongest protection against known fingerprinting techniques.
However, if one starts to deviate from this one and unique fingerprint, the user may end up being more visible and more easily trackable than with a standard browser like Chrome or Firefox.

\paragraph{Firefox's fingerprinting resistance}
Since its debut, the Tor Browser has been based on the \textit{Extended Support Release} (ESR) versions of Firefox.
With each new Firefox release, Tor developers had to update all their privacy-enhancing patches to continue building their browser.
In order to reduce as much as possible this time consuming process, the Tor Uplift project was launched in 2016~\cite{torUplift}.
Its goal is to bring most patches and security features of the Tor Browser directly into Firefox, including all the modifications made to counter fingerprinting.
Now, Tor developers do not have to redevelop their patches for Firefox anymore.
At the same time, Mozilla can experiment with advanced privacy features being tested in the Tor Browser to see if they could be brought to their audience.
From version 59 released in March 2018, a fingerprinting protection can be activated in Firefox with the \textit{privacy.resistFingerprinting} flag.
When it is enabled, the fingerprint is changed: the Canvas API is blocked and the user-agent, the timezone and the screen resolution are modified similarly to what the Tor Browser is doing.
The protection is being actively developed and the complete list of upcoming changes can be seen on ~\cite{firefoxProtection}.
Fingerprinting resistance is expected to get even stronger in the coming months with the launch of the Fusion project (Firefox USIng ONions - \cite{firefoxFusion}).

\subsubsection{UniGL}
Over the years, two studies have shown that the WebGL API can be used to create differences between devices by rendering complex 3D scenes~\cite{mowery12PixelPerfect,cao17cross}.
Wu et al. looked specifically at that API to identify the source of these differences~\cite{wu19webGL}.
Through manual testing and experiments, they found the single reason behind them: the results of floating-point operations can vary between devices across the various graphics layers of a system.
In order to make 3D rendering uniform, they designed a software solution called UniGL that redefines floating operations explicitly written in GLSL programs or implicitly invoked by WebGL.
That way, every device running UniGL will have the exact same WebGL fingerprint for a specific rendering task.

UniGL is a prime example of how uniformity can be achieved through a purely software solution that can be easily deployed to users.
By identifying the exact source of discrepancies, it is possible to remove differences without having to modify or update the hardware.

\subsection{Decreasing the surface of browser APIs}
\label{subsec:decreasing}

The last defense is to decrease the surface of browser APIs and reduce the quantity of information that can be collected by a tracking script.
One approach is to simply disable plugins so that additional fingerprinting vectors like Flash or Silverlight are not available to leak extra device information.

Another straight-forward way is to simply not run tracking scripts.
One can go into the browser preferences and disable the execution of JavaScript code for all web pages.
However, by doing so, the user will meet a static and broken web where it is impossible to login to most services.
An alternative is to use a browser extension like NoScript which uses a whitelist-based blocking approach 
~\cite{noscript}. 
By default, all JavaScript scripts are blocked and it is up to the user to choose which scripts can run.
The major problem with NoScript is that it is hard sometimes to distinguish which scripts are necessary to display a web page correctly and which domains belong to unwanted third parties. 
In the end, the user ends up authorizing all scripts on the page including the fingerprinting ones.

Another approach is to use ad and tracker blockers which block scripts and domains based on curated lists.
When a page is loaded, the extension analyses its content. 
If it finds a script or a domain that is present in one of its lists, it will block it.
The most popular addons based on this workflow are Adblock Plus~\cite{adblockplus}, Ghostery~\cite{ghostery}, uBlock Origin~\cite{ublockGithub} and Disconnect~\cite{disconnect}.
Merzdovnik et al. report on the effectiveness of these third-party tracker blockers on a large scale~\cite{merzdovnik2017block}.
One of the main downside of this type of extensions is that it can take a lot of time before a new script is detected and blocked, leaving the user vulnerable in the meantime.

Yu et al. proposed a concept in which users collectively identify unsafe data elements and report them to a central server~\cite{yu16cliqz}.
In their model, all data elements are considered unsafe when they are first reported.
Then, if enough users report the same value for a given script, the data elements are considered to be safe as it cannot be used to uniquely identify a user or a group of users.
Their approach is now in Cliqz~\cite{cliqz}.

The EFF who was behind the original Panopticlick study~\cite{eckersley2010Pano} released an extension called Privacy Badger in 2014~\cite{privacyBadger}.
The tool is similar to the approach chosen by Yu et al. to identify unsafe scripts but instead of relying on a global network of users, everything is computed locally by the extension. 
The list of blocked scripts is somehow unique to each instance of Privacy Badger as it is being built alongside the websites that the user visits.
However, the main downside of Privacy Badger is that the heuristic creation of blocking rules can be too aggressive and can lead to a high number of unresponsive websites as reported by~\cite{merzdovnik2017block}.

Sj\"{o}sten et al. designed a solution called Latex Gloves that addresses the problem of extension fingerprinting.
To protect against probing attacks, their solution restricts the accessibility of WARs through a whitelist system. 
When a request with the \textit{chrome-extension://} scheme is made, a browser extension checks if the domain was whitelisted and if not, blocks the request.
For revelation attacks, Latex Gloves rewrites the manifest file of extensions to specify on which domain it is allowed to run. This way, the browser would not reveal the presence of its extensions on a arbitrary page controlled by an attacker.
One of the downside of this approach is that whitelists require the user to maintain them and to allow each visited domain accordingly.

In terms of blocking protection, the last approach consists in disabling browser functions and even entire APIs to prevent trackers from using them.
This way, tracking scripts cannot collect values that could help them differentiate one device from another.
For example, an extension like CanvasBlocker~\cite{canvasBlocker} on Firefox disables the use of the Canvas API for all websites. 
The Tor Browser~\cite{torBrowser} blocks by default APIs like Canvas or WebGL and the Brave browser~\cite{brave} provides a built-in fingerprinting protection~\cite{braveProtection} against techniques like Canvas, WebGL, or AudioContext fingerprinting.

\subsection{Summary}
All in all, there is simply no ultimate solution against browser fingerprinting.
As this technique is anchored in years of web evolution, it cannot be fixed with a simple patch.
Changing the default behavior of the browser to fight it requires finding the right balance between privacy and usability and as we saw in this section, it can be very tricky.
One misstep and a protective solution can be rendered useless.

Datta et al. evaluated in depth 26 anti-fingerprinting tools~\cite{datta19antiFP} and came to the same conclusion: not all defense solutions are equal and some of them are performing better than others. 
For 24 of them, the protection they provide is apparently so marginal that it makes almost no difference not using them.
The authors also acknowledge that it is sometimes better to use one tool over another just because it is more popular even if it provides less protection.
The reason behind this is that it is better to hide in a large pool of users that have the same extension than being picked out as one of the few who uses this less popular one.

On the side of browser vendors, progress is being slowly made.
Since its debut, the Brave browser has had a built-in fingerprinting protection that blocks or disables several APIs~\cite{braveProtection}.
In 2018, Apple launched a version of Safari that specifically targets fingerprinting by removing differences between users~\cite{antiFPApple}.
Mozilla has launched in 2018 the Fusion project that is bringing modifications made in the Tor Browser directly in Firefox~\cite{firefoxFusion}.
They also turned on by default in 2019 a feature called Enhanced Tracking Protection that blocks fingerprinting scripts based on a list provided by Disconnect~\cite{antiFPMozilla}.
Finally, Google who currently has the biggest market share with Chrome~\cite{browserMarketShare} announced plans in August 2019 to ``aggressively block fingerprinting''~\cite{antiFPGoogle}. 
In the end, it is hard to assess the impact of all these defenses on browser fingerprinting going forward but the frantic pace at which the web keeps evolving will surely maintain the field alive and bring its load of surprises.

\section{Challenges in browser fingerprinting}
\label{sec:challenges}

In this section, we look at the usage of browser fingerprinting for two distinct purposes: 
tracking activities of the user and defending users from security threats. We them 
discuss questions and challenges surrounding the browser fingerprinting domain. 

\subsection{Usage of browser fingerprinting}
\label{sec:usage}

Though browser fingerprinting is often considered as a web tracking technology,
in practice is it used for a variety of purposes. 
We classify the usage of browser fingerprinting in two main categories:

\begin{itemize}
\item \textbf{Negative or destructive use} An unknown third party would want to track a user without her consent or to attack her device by identifying a known vulnerability.
\item \textbf{Positive or constructive use} Users can be warned if their device is out of date by recommending specific updates.
The security of online services can also be reinforced by verifying that a device is genuine and that it is known to the system. 
\end{itemize}

\subsubsection{Web Tracking}
\label{subsec:tracking}

As browser fingerprinting can uniquely identify a device on the web, the implications on privacy are important.
By collecting browser fingerprints on several websites, a third party can recognize a user and correlate his browsing activity within and across sessions.
Most importantly, the user has no control over the collection process as it is completely transparent since the tracking scripts are silent and executed in the background.
The Panopticlick study outlines in more details how fingerprinting can a be a threat to web privacy~\cite{eckersley2010Pano}.

\begin{itemize}
\item \textbf{Fingerprints as Global Identifiers} If a device has a fingerprint that is unique, it can be identified on the web without the need of other identifiers like a cookie or an IP address.
Peter Eckersley add in his study that it is ``akin to a cookie that cannot be deleted''.
Users funneling their network packets through a VPN (Virtual Private Network) are particularly vulnerable to browser fingerprinting as the VPN will only mask the IP address but it will not change the browser's information.

\item \textbf{Fingerprint + IP address as Cookie Regenerators}
Coupled with a fixed IP address, a browser fingerprint can be used to regenerate deleted cookies.
Researchers have already observed in the wild that any browser storage mechanisms like Flash local storage~\cite{soltani2010flash}, HTML5 Web storage~\cite{ayenson2011flash} or IndexedDB databases~\cite{acar14neverForgets}, can be used to ``respawn'' HTTP cookies.

\item \textbf{Fingerprint + IP address in the Absence of Cookies} In the absence of cookies, browser fingerprinting can be used to unmask different machines hiding behind the same IP address.
\end{itemize}

\subsubsection{Identifying device vulnerabilities}
\label{subsec:vulnerabilities}
A browser fingerprint is not just a simple collection of device-specific information.
It truly reflects the actual set of components that are running on a device.
By analysing its content, attackers can identify potential security vulnerabilities by cross-referencing the list of installed components with a database like CVE (Common Vulnerabilities and Exposures~\cite{cveWebsite}).
They can then design the perfect payload to target a specific device knowing its vulnerabilities in advance.
For example, through the \textit{navigator.plugins} property, one can know if a device is running an outdated version of the Flash plugin.
At the time of writing, the CVE database reports 1,045 Flash vulnerabilities and more than 84\% are labelled as critical, including the most recent ones~\cite{cveFlash}.
If the Flash player is not up to date, users open themselves to serious security risks as any attacker on the web could execute malicious code remotely on their device.

Launching a targeted attack with the help of browser fingerprinting is not new and has been observed in the wild.
Malwarebytes and GeoEdge have documented extensively with the ``Operation fingerprint'' how malicious advertising campaigns use fingerprinting to deliver malwares to vulnerable devices~\cite{operatinFingerprint}.
Their process is very straightforward.
They hide fingerprinting code directly into the JavaScript of fake advertisers and they look from there if the device is vulnerable or not.
If it is, the device will be presented with ``an ad laced with malicious code that ultimately redirects to an exploit kit''.
If it is not, the ad will be ``benign''.
To illustrate their findings, they detail several types of malvertising campaigns like the \textit{DoubleClick} or the \textit{musical4} campaigns.


\subsubsection{Patching vulnerable systems}
\label{subsec:patching}
Directly following the previous section, browser vulnerabilities could be identified with the aim of patching them.
In 2015, Duo Security reported that 46\% of corporate PCs ran outdated versions of browsers, Flash and Java~\cite{duoVulnerabilities}.
With a simple security scan, system administrators who handle thousands of different configurations on a network could easily identify devices with outdated components and they could deploy fixes and updates really quickly.

\subsubsection{Bot and fraud prevention}
\label{subsec:bot-fraud}
Another use of browser fingerprinting is to improve security on the web by verifying the actual content of a fingerprint.
As there are many dependencies between collected attributes, it is possible to check if a fingerprint has been tampered with or if it matches the device it is supposedly belonging to (see Section~\ref{subsec:increasing} for details on the inconsistency problem).

ThreatMetrix, a security company that specializes in the verification of online transactions, announced in 2010 the adoption of browser fingerprinting techniques to prevent online fraud~\cite{FPthreatmetrix}.
They wrote that fraudsters change their IP address, delete cookies and botnet scripts randomize device attributes.
Moreover, relying exclusively on cookies is no longer adequate to verify an online transaction.
Other security companies like Distil Networks~\cite{FPdistil}, MaxMind~\cite{FPmaxmind}, PerimeterX~\cite{FPPermiter}, IPQualityScore~\cite{FPQuality}, ShieldSquare~\cite{FPShieldSquare} or Sift Science~\cite{FPSift} also utilize browser fingerprinting to detect bots and unusual activity.
In that landscape, companies are turning to browser fingerprinting to be competitive in this continual arms race against fraudsters.

On the academic side, the literature on fraud detection is much thinner with only a single publication addressing this problem.
Researchers at Google designed a solution called Picasso based on canvas fingerprinting to filter inorganic traffic~\cite{bursztein2016Picasso}.
By using specific graphical primitives from the canvas API, they are able to successfully detect the browser and OS family of a device and see if there is a mismatch between the exhibited fingerprint and the actual device running the code.
For example, they can distinguish between traffic sent by an authentic iPhone running Safari on iOS from an emulator or desktop client spoofing the same configuration.
They add that the applications are numerous including locking non-mobile clients from application marketplaces, detecting rogue login attempts and identifying emulated clients.
Their study does not give information on the deployment of Picasso in a current Google solution but a talk at Black Hat Asia 2016 hints at its integration into Google's reCAPTCHA technology~\cite{sivakornm16recaptcha}.

\subsubsection{Augmented authentication}
\label{subsec:authentication}
At a time where passwords are the go-to solution for authentication on the web, browser fingerprinting can provide a much needed addition to reinforce the security of online accounts.
By verifying the fingerprint of a device at login time, a system can easily block unauthorized access from new and unknown devices.
Alaca et al. studied extensively the use of device fingerprinting for web authentication~\cite{alaca2016Authentication}. 
They classify in total 29 different attributes from browser information to the network stack according to criteria like repeatability, low resource use or spoofing resistance.
One important aspect considered in their study is the notion of stability.
As a fingerprint is the direct reflection of what is installed on a device, a browser fingerprint constantly changes. 
It is then up to the login system to decide if the differences between two fingerprints are acceptable or not.
For example, does a change of browser version in the user-agent come from a legitimate update of the device or from a different device altogether?
If ten fonts are removed, did the user uninstall a particular software or does it come from a different device that does not belong to the user?
These questions have no easy answer and each collected attribute has its own behavior depending on the system being used or the type of the device.

Spooren et al. looked at mobile devices and noticed that mobile fingerprints are predictable contrarily to desktop fingerprints~\cite{spooren2015MobileFPHarmful}.
The same authors also investigated the use of battery information for mobile devices in a multi-factor authentication scheme~\cite{spoorenPJ17}.
By using binary classifiers to classify battery draining and charging behaviors, they confirm that battery charge measurements can be used to contribute to an active authentication system.

In 2019, two different studies reported on the use of canvas fingerprinting to augment authentication with a challenge-response scheme.
After the user connects with a password, the browser is asked to render a canvas image that will then be verified by the server.
While they use a very similar protocol, the inner-workings of the two systems are very different.
The first by Rochet et al. uses deep learning to create a personalized model for each device~\cite{rochet19swat}. 
After a training phase with 2,000 canvas images, features are extracted to build a binary classification model for the user.
To accept or deny connection, the browser must score above a specific threshold and the authors discuss different strategies in the paper for setting that particular threshold.
The second by Laperdrix et al. adopts a different approach~\cite{laperdrix19Morellian}.
To verify the current connection, the browser must be able to replicate the exact same canvas rendering that was observed in the previous connection. 
If the current image matches with pixel-perfect precision the previous one, the device is allowed to continue. 
If not, it is blocked and the user must authenticate through another means.
In the end, the authors of both papers state clearly that their solution cannot be the only means of authentication as it should be used in a multi-factor authentication scheme.
They both describe the advantages and limits of their solution and they highlight how the level of security can be changed depending on the strategy adopted with each tool.

Finally, some companies include in their portfolio fingerprinting solutions to augment authentication.
SecurAuth is a provider of an adaptive access control solution. 
As part of their multi-factor authentication process, they include an heuristic-based authentication system through device fingerprinting~\cite{securauth}.
Another company called Iovation has a solution named ClearKey~\cite{clearkey} that integrates the collection of device information as part of their multi-factor authentication framework.
They provide their own approach to deal with fingerprint changes with fuzzy logic algorithms~\cite{clearkeySheet}.

All in all, the amount of research done to use browser fingerprinting positively is extremely thin and we hope that this area will see major advancements in the future.

\subsection{Current challenges}

We now discuss the current challenges in browser fingerprinting in both research and industry. 

\subsubsection{Arms-race between new fingerprinting methods and protection mechanisms}

Browser fingerprinting should be placed in the larger debate about online tracking.
If third-parties feel the need to resort to such kinds of techniques to bypass current protection mechanisms, the current ad ecosystem is not quite right.
On one side, ad blockers are seeing a surge in popularity~\cite{pageFairReport} and companies like Mozilla~\cite{firefoxProtection} or Brave~\cite{braveProtection} are integrating a native fingerprinting protection directly in their browsers.
On the other side, major web actors are struggling to decide what is the right path going forward as illustrated by the divide created by two new ad initiatives. 
One called \textit{Coalition for Better Ads}~\cite{initiative1} is supported by actors like Google, Facebook, Microsoft or Criteo.
The other one called \textit{Acceptable Ads} initiated by Eyeo GmbH, developer of the famous Adblock Plus extension, is led by a coalition of digital rights organizations, advertising agencies, content creators and academics~\cite{initiative2}.
In the end, there is no clear direction of where the industry is going. 
Browser fingerprinting is caught in the crossfire of this ongoing debate as developers are already adding defensive solutions in anticipation of what could be and the consequence is that it affects all companies from the biggest ones to the smallest.


Since browser fingerprinting can pose a serious threat to privacy by bypassing current protection mechanisms, 
{\em should researchers study it?}
We believe the answer to this question has two sides.
Indeed, researching browser fingerprinting is needed because we want to inform users, developers, policy makers and lawyers about it so that they can make informed decision going forward. 
By knowing its underlying mechanisms, we can understand what is possible with it and act accordingly.
We can also design appropriate defenses and protect ourselves against it.
For example, the web crawl performed by Englehardt et al.~\cite{englehardt2016OpenWPM} discovered the existence of AudioContext fingerprinting in the wild and the research into the Battery Status API~\cite{olejnik15battery} revealed the threat that was hidden in our browsers for several years.
Thanks to these studies, fingerprinting defenses made a step forward to improve online privacy for users.
However, on the other side, we do not want to promote aggressive use of this technology and make it more efficient.
Research in offensive security is encouraged today so that an attack is known to the public and not kept in the hands of a few adversaries.
The same reasoning applies here for browser fingerprinting where studying it benefits users more than it benefits attackers.


\subsubsection{How to effectively measure uniqueness of fingerprints?}
An important challenge in browser fingerprinting is to understand exactly in which circumstances 
fingerprinting is effective 
in uniquely recognizing website visitors.
While Panopticlick~\cite{eckersley2010Pano} and AmIUnique~\cite{laperdrix16Beauty} showed a high rate of 
fingerprint uniqueness, the Hiding in the Crowd study~\cite{gomez18hiding} highlighted the potential demographic 
issues in browser fingerprinting research.
Depending on the targeted audience and the type of users who connects to a website, the capacity to 
uniquely identify users with only their fingerprint can greatly vary.
Moreover, only one study has partially studied the correlation between fingerprint uniqueness and the size of a dataset in 
case of fingerprinting via browser extensions~\cite{Guly-etal-18-WPES}.

Predicting uniqueness for large datasets can be a successful process as reported by \cite{achara15Smartphone} 
but it remains to be seen if such  an approach can work for fingerprints, especially because of their constantly changing nature. 
This question is also impacted by the quality of current fingerprint datasets. 
As it is complicated for researchers to collect data on a large scale, their experiments can run for several months up to more than a year.
In that time frame, browsers are updated and some APIs may undergo some key changes.
Comparing two fingerprints that were collected several months apart does not make sense for identification or tracking. 
Additional research in this area is definitely needed as the reality of fingerprint tracking may be much more nuanced than what is currently reported.

\subsubsection{How to detect fingerprinting and measure its adoption?}

The next 
challenge is to measure precisely 
adoption of browser fingerprinting online. 
In order to quantify the number of websites that are currently using fingerprinting scripts on the Internet, one needs the means to identify them.
However, even if the collection process in a browser is straightforward, the reality is in fact much more complex.
If a script accesses the language header and the platform, it could be for legitimate purposes to tailor the current web page to the user's device and propose the right version of a software to a user.
But it could also be the first-step towards building a complete browser fingerprint.
If a script makes hundreds of calls to the WebGL API, it may be rendering a 3D animation in the browser. 
But it may also test complex animations to differentiate this device from others.
These simple examples illustrate that the line between a benign script and a fingerprinting one is far from being clearly defined.
When crawling the web, researchers are facing a lot of challenges to classify scripts correctly as the goal of two scripts can vastly vary even if they present very similar content.
Dynamic information flow analysis is required here to precisely identify fingerprinting scripts. 
We discuss this challenge in more details in Section~\ref{subsec:watchers}.

\subsubsection{A constant need to monitor for security and privacy issues of new browser APIs}

One of the main challenges surrounding browser fingerprinting is that it is hard to assess precisely what is possible with it.
As its mechanisms are entirely defined by current web browser technologies, its exact contours are constantly changing.
Each new browser version that adds, modifies or even removes an API has a direct impact on the domain.
For example, the introduction of the Canvas API brought new capabilities to the domain while the end of NPAPI plugins removed a strong source of information. 
The browser fingerprinting of the past is already different from the ones we see today and will surely be different from the one we will encounter in the next few years.
Browser vendors and standard organizations are continually shaping the future of the domain as they discuss about what is next for the web and what will run in tomorrow's browsers.
One certainty is that they will be very careful going forward when designing new APIs and releasing them to the public.
The Battery Status API showed that privacy cannot be an afterthought and that extra help from the industry and the research worlds is needed to fully asses any potential lurking threats.

%

\subsubsection{How to detect violations of data protection laws?}


As a whole, browser fingerprinting can be particularly dangerous to privacy as browsers do not provide any controls over it.
In Europe, 
the European Data Protection Board (EDPB)\footnote{European Data Protection Board (EDPB) is used to be called Article 29 Data Protection Working Party before 2018.}, which seeks to harmonize the application of data protection 
rules throughout the EU, published back in 2014 an opinion on device fingerprinting~\cite{EUFPDirective} 
and in 2012 an opinion on exemptions from user consent~\cite{WP29-04-2012}.  
According to these two opinions, what determines user consent in case of browser fingerprinting is 
{\em the purpose of usage}. For example, fingerprinting can be used without consent for user 
authentication~\cite[Section 3.2]{WP29-04-2012}, that corresponds to the usage we described in 
Section~\ref{subsec:authentication}. 
Consent is also not required for ``increasing the security of the service that has been explicitly requested by the user''~\cite[Section 3.3]{WP29-04-2012} (see our Sections~\ref{subsec:patching} and \ref{subsec:bot-fraud} for more details). 
It is also clear that browser fingerprinting used for tracking  (see Section~\ref{subsec:tracking}) or profiling 
(Section~\ref{subsec:vulnerabilities}) requires user consent. Interestingly, our intuitive classification of 
negative and positive use of fingerprinting set in Section~\ref{sec:usage} is 
closely related to the requirements of user consent described by the European Data Protection Board (EDPB).  

Next to the General Data Protection Regulation (GDPR)~\cite{GDPR} that came in force on 
May 25, 2018, another law that specifically regulates web tracking in the EU 
is currently being updated to 
\emph{ePrivacy Regulation}~\cite{ePrivacyEC,ePrivacyEP}. 
The latest amendments to the ePrivacy Regulation draft~\cite{ePrivacyEP}  
requires 
user's consent for fingerprinting ``and for specific and transparent purposes'', 
but with some exceptions. 
%
%
For example, 
it allows websites to check browser's configuration for any needed security updates without user's consent.
It also allows first-party servers to use fingerprinting (and any stateful tracking) for web audience measurement but 
requires that the user still maintains the right to object, and that no personal data is made accessible to any third party.
%
%
However, EDBP
latest report~\cite{WP292017} is 
very specific: the analytics technology used on a website
should prevent re-identification, and the collected data cannot be linked in any way to other identifiable data.
Under these requirements, it is questionable whether fingerprinting can be used for audience measurement at all
without user's consent as it is possible to re-identify users based on their fingerprints.	


In the end, regulators are already in an uphill battle to verify if companies are complying with these European rules as the necessary controls cannot easily be performed online. 
For fingerprinting, the detection is 
very complex as the browser has no mechanism dedicated to detecting it precisely. 
As a fingerprint is an extremely versatile object, it is hard to reason about it and 
to verify that a website is partaking in fingerprinting activities or not. 
Moreover, it is even a greater challenge from a legal perspective to obtain {\em the purpose} of the 
usage of browser fingerprinting in order to establish whether user consent is needed. 
Regulators will need to find new ways to cooperate with companies to make sure that the privacy of 
users is respected.

\section{Conclusion}
\label{sec:conclusion}

The development of the Internet along with progress in mobile technology brought a booming diversity of devices at the turn of the century.
This diversity gave birth to browser fingerprinting, a simple technique that consists in collecting information about the configuration and the composition of a user's device.
Its fascinating aspect is that it is at a crossroads between companies, academic research groups, law makers and privacy advocates.
As it got out of the research lab, it has a concrete impact on the web as it is now used in real-world scenarios.
For business companies, browser fingerprinting represents an alternative to current methods of tracking and identification at a time where the ad landscape is undergoing tremendous changes with the rise of ad blockers.
For research groups, browser fingerprinting brought unexpected questions about the privacy status of current and future web APIs.
Especially, the work done by researchers on the Battery API exposed the possible privacy problems that could be hiding in current browsers.
For law makers, browser fingerprinting represents an additional tracking mechanism that must be regulated so that the control is given back in users' hands. 
For journalists, activists, businesses or members of the military that rely on the confidentiality and privacy of their communications, they must now take this technique into account to protect their activities.
All in all, browser fingerprinting is still a fairly new technique.
Yet, it already had a lot of impact in its short time of existence.
Our effort to systematize existing knowledge proves there are still many open challenges and problems to be solved as researchers and developers are coming to grasp with its intricacies.
We hope that our paper will provide the necessary basis for researchers to analyze even further the inner-workings of fingerprinting as novel solutions based on it have the potential to provide real-world benefits to millions by improving online security.





\bibliographystyle{ACM-Reference-Format}
\bibliography{bib/articles.bib,bib/url.bib}

\end{document}